\RequirePackage{ifpdf}
\ifpdf % We are running pdfTeX in pdf mode
\documentclass[pdftex]{sigma}
\else
\documentclass{sigma}
\fi

\newcommand{\rd}{\partial}

\newcommand{\CC}{\mathbf{C}}
\newcommand{\PP}{\mathbf{P}}

\newcommand{\res}{\mathop{\mathrm{res}}}
\newcommand{\pbar}{\bar{p}}
\newcommand{\qbar}{\bar{q}}
\newcommand{\tbar}{\bar{t}}
\newcommand{\ubar}{\bar{u}}

\newcommand{\zbar}{\bar{z}}
\newcommand{\Bbar}{\bar{B}}
\newcommand{\Fbar}{\bar{F}}
\newcommand{\Gbar}{\bar{G}}
\newcommand{\Lbar}{\bar{L}}
\newcommand{\Vbar}{\bar{V}}
\newcommand{\bst}{\boldsymbol{t}}

\newcommand{\bszero}{\boldsymbol{0}}
\newcommand{\bstbar}{\bar{\bst}}
\newcommand{\htilde}{\tilde{h}}
\newcommand{\Mtilde}{\tilde{M}}

\newcommand{\hhat}{\hat{h}}
\newcommand{\betahat}{\hat{\beta}}

\numberwithin{equation}{section}

\begin{document}

\allowdisplaybreaks

\renewcommand{\thefootnote}{$\star$}

\renewcommand{\PaperNumber}{102}

\FirstPageHeading

\ShortArticleName{Old and New Reductions of Dispersionless Toda Hierarchy}

\ArticleName{Old and New Reductions\\ of Dispersionless Toda Hierarchy\footnote{This
paper is a contribution to the Special Issue ``Geometrical Methods in Mathematical Physics''. The full collection is available at \href{http://www.emis.de/journals/SIGMA/GMMP2012.html}{http://www.emis.de/journals/SIGMA/GMMP2012.html}}}

\Author{Kanehisa TAKASAKI}

\AuthorNameForHeading{K.~Takasaki}

\Address{Graduate School of Human and Environmental Studies,
Kyoto University,\\
Yoshida, Sakyo, Kyoto, 606-8501, Japan}
\Email{\href{mailto:takasaki@math.h.kyoto-u.ac.jp}{takasaki@math.h.kyoto-u.ac.jp}}
\URLaddress{\url{http://www.math.h.kyoto-u.ac.jp/~takasaki/}}

\ArticleDates{Received June 06, 2012, in f\/inal form December 15, 2012; Published online December 19, 2012}

\Abstract{This paper is focused on geometric aspects of two particular types
of f\/inite-variable reductions in the dispersionless Toda hierarchy.
The reductions are formulated in terms of ``Landau--Ginzburg potentials''
that play the role of reduced Lax functions.  One of them is
a generalization of Dubrovin and Zhang's trigonometric polynomial.
The other is a transcendental function, the logarithm of which resembles
the waterbag models of the dispersionless KP hierarchy.
They both satisfy a radial version of the L\"owner equations.
Consistency of these L\"owner equations yields a radial version
of the Gibbons--Tsarev equations.  These equations are used
to formulate hodograph solutions of the reduced hierarchy.
Geometric aspects of the Gibbons--Tsarev equations are explained
in the language of classical dif\/ferential geometry
(Darboux equations, Egorov metrics and Combescure transformations).
Flat coordinates of the underlying Egorov metrics are presented.}


\Classification{35Q99; 37K10; 53B50; 53D45}

\vspace{-2mm}

\renewcommand{\thefootnote}{\arabic{footnote}}
\setcounter{footnote}{0}

\section{Introduction}

\looseness=-1
The notion of f\/inite-variable reductions
in dispersionless integrable hierarchies \cite{Krichever94}
has rich geometric contents that range
from the classical dif\/ferential geometry
of orthogonal curvilinear coordinates \cite{Darboux-oclc}
to the modern theory of Frobenius manifolds \cite{Dubrovin-2dtft}.
Moreover, the L\"owner equations \cite{Lowner23},
f\/irst introduced to solve the Bieberbach conjecture
in the univalent function theory, also play
a fundamental role in this issue.  In this paper,
we consider the dispersionless Toda hierarchy \cite{TT95}.
Although a general scheme of f\/inite-variable reductions
in this case is already established \cite{CLR09, Manas04,TT06,TTZ06},
f\/inding interesting examples is another issue.
We here report two examples, one being a generalization
of an ``old'' example, and the other being a ``new'' one.
These examples turn out to f\/it well into
the aforementioned geometric perspectives.
We believe that these examples have their own
interesting features, part of which will be presented later on.

As in the case of other dispersionless integrable systems
\cite{Krichever94}, a f\/inite-variable reduction
of the dispersionless Toda hierarchy can be characterized
by a globally def\/ined reduced Lax function~$\lambda(p)$.
Borrowing the terminology of topological f\/ield theories \cite{Dijkgraaf92},
let us call this reduced Lax function a ``Landau--Ginzburg potential''\footnote{This is somewhat problematical, because
there is no guarantee that the reduced Lax function
has an associated topological f\/ield theory or
a Frobenius manifold.}.
The two examples addressed in this paper have
Landau--Ginzburg potentials of the following form:

{\bf Case I:}
\begin{gather}
  \lambda(p) = p^{-N}\prod_{i=1}^M(p - b_i)^{\kappa_i},\qquad
  \text{where}\quad \sum_{i=1}^M \kappa_i - N > 0 , \quad
 \kappa_i \not= 0 ,  \quad N \not= 0.
  \label{lambda(I)}
\end{gather}

{\bf Case~II:}
\begin{gather}
  \lambda(p) = \prod_{i=1}^M(p - b_i)^{\kappa_i}
               \exp\left(\sum_{k=1}^N c_kp^{-k}\right),\qquad
  \text{where} \quad \sum_{i=1}^M \kappa_i > 0, \quad
 \kappa_i \not= 0 , \quad N > 0.
  \label{lambda(II)}
\end{gather}

(\ref{lambda(I)}) is a generalization of the well known
trigonometric Landau--Ginzburg potential studied
by Dubrovin and Zhang \cite{DZ98}.
Dubrovin and Zhang's Landau--Ginzburg potential
corresponds to the case where $N > 0$ and
$\kappa_1 = \cdots = \kappa_M =1$.  This case contains,
as particular examples, the dispersionless limit
of the 1D Toda and bigraded Toda hierarchies \cite{Carlet06}.
By allowing~$\kappa_i$'s and~$N$ to take negative values,
we can include therein, for example, the dispersionless limit
of the Ablowitz--Ladik hierarchy~\cite{AL75} and its possible
generalizations as well.  Thus this apparently ``old'' case
itself deserves to be studied in detail.

(\ref{lambda(II)}) is presumably a new example
that has never been studied in the literature.
This Landau--Ginzburg potential is not a rational function
of~$p$.  In this respect, (\ref{lambda(II)}) is
conceptually similar to the so called ``waterbag'' models
that, too, have irrational Landau--Ginzburg potentials.
The waterbag models were f\/irst presented
in the celebrated work of Gibbons and Tsarev \cite{GT99, GT96}
on reductions of the Benney hierarchy,
and have been further studied in the dispersionless
KP hierarchy from various points of view
\cite{BK04,Chang06,Chang07dkp,FS06,Pavlov06}.
In the most general formulation presented
by Ferguson and Strachan~\cite{FS06},
the Landau--Ginzburg potential is a sum of a polynomial
and logarithmic terms of the following form
\begin{gather}
  \lambda(p) = p^N + \sum_{k=2}^N c_kp^{N-2}
    + \sum_{i=1}^M\kappa_i\log(p - b_i).
\label{FS-lambda}
\end{gather}
Analogy with the waterbag models becomes more manifest
in the logarithmic form
\begin{gather}
  \log\lambda(p)
  = \sum_{i=1}^M \kappa_i\log(p-b_i)
    + \sum_{k=1}^N c_kp^{-k}
\label{log-lambda(II)}
\end{gather}
of~(\ref{lambda(II)}), which takes almost the same form
as~(\ref{FS-lambda}) except that the (Laurent) polynomial part
of the latter is a polynomial in $p^{-1}$.\footnote{Another delicate dif\/ference is that
the polynomial part of (\ref{FS-lambda}) is monic
(namely, the coef\/f\/icient of the highest degree term is~1)
and has no next-highest degree term.  In contrast,
the (Laurent) polynomial part of~(\ref{log-lambda(II)})
is not monic, and has no constant term.}
In this sense, one may think of~(\ref{lambda(II)})
as a variation of the waterbag models.

Let us mention that there are a few proposals
of waterbag models for the dispersionless Toda hierarchy.
The earliest one is Yu's model \cite{Yu00}\footnote{Actually, Yu considered a dispersionless limit
of the discrete KP hierarchy.  This hierarchy is a subsystem
of the dispersionless Toda hierarchy, and one can readily
translate Yu's waterbag model to the language of the latter.}.
Seemingly unaware of Yu's work, Chang proposed
three types of waterbag models~\cite{Chang07dtoda}.
(\ref{lambda(I)})~is a slight generalization
of Chang's f\/irst model (which amounts to the case
where $N = -1$, and for which Chang considered
a Frobenius structure).  Chang's second and third models
are actually a generalization of Yu's model.
The logarithmic expression~(\ref{log-lambda(II)})
of our second case resembles Chang's second and third models.
We, however, feel that the exponentiated form~(\ref{lambda(II)}) is more natural
than the logarithmic expression.

This paper is organized as follows.
Section~\ref{section2} is devoted to the Lax formalism.
We show, by a direct method, that~(\ref{lambda(I)}) and~(\ref{lambda(II)})
give consistent reductions of the Lax equations
of the dispersionless Toda hierarchy.
The two-variable cases turn out to contain some well known
(as well as new) dispersionless integrable hierarchies.
Section~\ref{section3} is focused on the L\"owner equations
and the associated Gibbons--Tsarev equations.
The L\"owner equations relevant to the dispersionless
Toda hierarchy are a ``radial'' version of
the ``chordal'' L\"owner equations
in the aforementioned work of Gibbons and Tsarev \cite{GT99, GT96}.
We show that~(\ref{lambda(I)}) and~(\ref{lambda(II)})
satisfy the radial L\"owner equations, and use these equations
to formulate the generalized hodograph method
\cite{Tsarev93, Tsarev85} for these reductions.
Section~\ref{section4} presents some geometric implications of
the Gibbons--Tsarev equations in the language of
Darboux equations, Egorov metrics and Combescure transformations.
We show that three particular Egorov metrics underlie
the radial Gibbons--Tsarev equations.
Section~\ref{section5} deals with f\/lat coordinates of the Egorov metrics.
Results in this section is somewhat restrictive.
We encounter dif\/f\/iculties in generalizing the dual pair
of Frobenius structures of Dubrovin and Zhang \cite{DZ98}
to the more general Landau--Ginzburg potential~(\ref{lambda(I)}).
Moreover, we can construct only a single Frobenius structure
without a dual for the Landau--Ginzburg potential~(\ref{lambda(II)}).  The method and the result for the latter,
however, exhibits remarkable similarities
with the work of Ferguson and Strachan~\cite{FS06}.

\section{Lax equations}\label{section2}

\subsection{Lax formalism of dispersionless Toda hierarchy}\label{section2.1}

The Lax equations of the dispersionless Toda hierarchy \cite{TT95}
are formulated by two Lax func\-tions\footnote{$z$ and $\zbar$ amount to $\mathcal{L}$
and $\mathcal{\Lbar}^{-1}$ in our previous notations \cite{TT95}.}~$z(p)$,~$\zbar(p)$
of a spatial variable $s$, a momentum variable $p$,
and two sets of time variables $\bst = (t_1,t_2,\ldots)$
and $\bstbar = (\tbar_1,\tbar_2,\ldots)$.\footnote{Throughout this paper, the overline ``$\bar{\quad}$''
does not mean complex conjugation.  For example,
$t_n$ and $\tbar_n$ are independent variables.}
$p$ is a ``classical limit'' of the shift operator $e^{\rd/\rd s}$
that satisf\/ies the twisted canonical commutation relation
\[
  \big[e^{\rd/\rd s},  s\big] = e^{\rd/\rd s}.
\]
In the classical (or ``long-wave'') limit,
this commutation relation turns into
the Poisson commutation relation
\[
  \{p,s\} = p.
\]
This Poisson bracket can be extended to arbitrary functions
of $s$ and $p$ as
\[
  \{f,g\} = p\left(\frac{\rd f}{\rd p}\frac{\rd g}{\rd s}
          - \frac{\rd f}{\rd s}\frac{\rd g}{\rd p}\right).
\]

In the most general formulation, the Lax functions
$z(p)$ and $\zbar(p)$ are understood to be
mutually independent formal Laurent series of $p$
of the form
\begin{gather*}
  z(p) = p + u_1 + u_2p^{-1} + \cdots, \qquad
  \zbar(p) = \ubar_0 p^{-1}  + \ubar_1 + \ubar_2p + \cdots.
\end{gather*}
The coef\/f\/icients $u_n = u_n(s,\bst,\bar{\bst})$
and $\ubar_n = \ubar_n(s,\bst,\bar{\bst})$ are dynamical variables.
The leading coef\/f\/icient $\ubar_0$ is assumed to take
the exponential form
\begin{gather*}
  \ubar_0 = e^{\phi},\qquad \phi = \phi(s,\bst,\bstbar).
\end{gather*}
Let us def\/ine the polynomials $B_n(p)$ and $\Bbar_n(p)$
in $p$, $p^{-1}$ as
\begin{gather*}
  B_n(p) = (z(p)^n)_{\ge 0}, \qquad
  \Bbar_n(p) = (\zbar(p)^n)_{<0},
\end{gather*}
where $( \ )_{\ge 0}$ and $( \ )_{<0}$
are projection operators acting on the linear space
of Laurent series as
\begin{gather*}
  \left(\sum_{n=-\infty}^\infty a_np^n\right)_{\ge 0}
  = \sum_{n\ge 0}a_np^n,\qquad
  \left(\sum_{n=-\infty}^\infty a_np^n\right)_{<0}
  = \sum_{n<0}a_np^n.
\end{gather*}
Time evolutions are generated by the Lax equations
\begin{gather}
  \frac{\rd z(p)}{\rd t_n} = \{B_n(p), z(p)\},\qquad
  \frac{\rd z(p)}{\rd\tbar_n} = \{\Bbar_n(p), z(p)\},\nonumber\\
  \frac{\rd\zbar(p)}{\rd t_n} = \{B_n(p),\zbar(p)\},\qquad
  \frac{\rd\zbar(p)}{\rd\tbar_n} = \{\Bbar_n(p),\zbar(p)\}.  \label{dToda-Lax-eq}
\end{gather}
It is convenient to introduce the complementary generators
\begin{gather*}
  B^c_n(p) = (z(p)^n)_{<0},\qquad
  \Bbar^c_n(p) = (\zbar(p)^n)_{\ge 0}
\end{gather*}
as well.  The Lax equations can be thereby rewritten as
\begin{gather}
  \frac{\rd z(p)}{\rd t_n} = \{z(p), B^c_n(p)\},\qquad
  \frac{\rd z(p)}{\rd\tbar_n} = \{z(p), \Bbar^c_n(p)\},\nonumber\\
  \frac{\rd\zbar(p)}{\rd t_n} = \{\zbar(p), B^c_n(p)\},\qquad
  \frac{\rd\zbar(p)}{\rd\tbar_n} = \{\zbar(p), \Bbar^c_n(p)\}. \label{dToda-dualLax-eq}
\end{gather}

\subsection[Landau-Ginzburg potential as reduced Lax function]{Landau--Ginzburg potential as reduced Lax function}\label{section2.2}

We now specialize the Lax equations to the case
where the formal (or local) Lax functions
$z(p)$ and $\zbar(p)$ are linked with the globally def\/ined
Landau--Ginzburg potential $\lambda(p)$ as follows:
\begin{itemize}\itemsep=0pt
\item[(I)] For the Landau--Ginzburg potential (\ref{lambda(I)}),
\begin{gather*}
  z(p) = \lambda(p)^{1/\Mtilde} \quad\text{as $p \to \infty$},\qquad
  \zbar(p) = \lambda(p)^{1/N} \quad\text{as $p \to 0$},%\label{zzbar-lambda(I)}
\end{gather*}
where $\Mtilde = \sum\limits_{i=1}^M\kappa_i - N$.
Recall that $\Mtilde$ is assumed to be positive.
$N$ can be both positive and negative.
\item[(II)] For the Landau--Ginzburg potential (\ref{lambda(II)}),
\begin{gather*}
  z(p) = \lambda(p)^{1/\Mtilde} \quad\text{as $p \to \infty$},\qquad
  \zbar(p) = \left(\log\lambda(p)\right)^{1/N} \quad\text{as $p \to 0$},%\label{zzbar-lambda(II)}
\end{gather*}
where
$\Mtilde = \sum\limits_{i=1}^M\kappa_i$.
Recall that $\Mtilde$ and $N$ is assumed to be positive.
\end{itemize}
The Lax equations (\ref{dToda-Lax-eq}) and
its complementary form (\ref{dToda-dualLax-eq})
thus reduce to the Lax equations
\begin{gather}
  \frac{\rd\lambda(p)}{\rd t_n} = \{B_n(p), \lambda(p)\}
                                = \{\lambda(p), B^c_n(p)\},\nonumber\\
  \frac{\rd\lambda(p)}{\rd\tbar_n} = \{\Bbar_n(p), \lambda(p)\}
                              = \{\lambda(p), \Bbar^c_n(p)\}\label{lambda-Lax-eq}
\end{gather}
for $\lambda(p)$.  We can conf\/irm that
this is a consistent reduction procedure in the following sense.

\begin{theorem}\label{thm:reduced-Lax-eq}
The Lax equations \eqref{lambda-Lax-eq} are equivalent
to a system of first order evolutionary equations
of the form
\begin{gather*}
  \frac{\rd b_i}{\rd t_n} = F_{in},\qquad
  \frac{\rd b_i}{\rd\tbar_n} = \Fbar_{in}
\end{gather*}
for the Landau--Ginzburg potential \eqref{lambda(I)}
and
\begin{gather*}
  \frac{\rd b_i}{\rd t_n} = F_{in},\qquad
  \frac{\rd c_k}{\rd t_n} = G_{kn},\qquad
  \frac{\rd b_i}{\rd\tbar_n} = \Fbar_{in},\qquad
  \frac{\rd c_k}{\rd\tbar_n} = \Gbar_{kn}
\end{gather*}
for the Landau--Ginzburg potential \eqref{lambda(II)},
where $F_{in}$, $\Fbar_{in}$, etc.\
are functions of $b_1,\ldots,b_M$, $c_1,$ $\ldots,c_N$
and their first order derivatives with respect to~$s$.
\end{theorem}

\begin{proof}
Since the two cases can be treated in the same way,
let us consider the Case~II only.
We can rewrite the Lax equations~(\ref{lambda-Lax-eq})
in terms of $\log\lambda(p)$ as
\begin{gather*}
  \frac{\rd\log\lambda(p)}{\rd t_n}
    = \{B_n(p), \log\lambda(p)\}
    = \{\log\lambda(p), B^c_n(p)\},\\
  \frac{\rd\log\lambda(p)}{\rd\tbar_n}
    = \{\Bbar_n(p), \log\lambda(p)\}
    = \{\log\lambda(p), \Bbar^c_n(p)\}.
\end{gather*}
The left hand side are linear combinations
of $1/(p-b_i)$'s and $p^{-k}$'s:
\begin{gather*}
  \frac{\rd\log\lambda(p)}{\rd t_n}
  = - \sum_{i=1}^M \frac{\rd b_i}{\rd t_n}\frac{\kappa_i}{p-b_i}
    + \sum_{k=1}^N\frac{\rd c_k}{\rd t_n}p^{-k},\\
  \frac{\rd\log\lambda(p)}{\rd\tbar_n}
  = - \sum_{i=1}^M \frac{\rd b_i}{\rd\tbar_n}\frac{\kappa_i}{p-b_i}
    + \sum_{k=1}^N\frac{\rd c_k}{\rd\tbar_n}p^{-k}.
\end{gather*}
Since $B_n(p)$ and $\Bbar_n(p)$ are polynomials
in~$p$ and~$p^{-1}$, the right hand sides
are rational functions of~$p$.  Actually,
because of the two complementary expressions,
they turn out to be linear combinations of~$1/(p-b_i)$'s
and~$p^{-k}$'s.  For example, $\{B_n(p),\log\lambda(p)\}$
can be expanded as
\begin{gather*}
  \{B_n(p),\log\lambda(p)\}
   = \sum_{i=1}^M\{B_n(p),\kappa_i\log(p-b_i)\}
     + \sum_{k=1}^N\big\{B_n(p),c_kp^{-k}\big\} \\
\hphantom{\{B_n(p),\log\lambda(p)\}}{}
= \sum_{i=1}^M p\left(
        - \frac{\rd B_n(p)}{\rd p}
          \frac{\rd b_j}{\rd s}\frac{\kappa_i}{p-b_i}
        - \frac{\rd B_n(p)}{\rd s}\frac{\kappa_i}{p-b_i}\right) \\
\hphantom{\{B_n(p),\log\lambda(p)\}=}{}
 + \sum_{k=1}^N p\left(
        \frac{\rd B_n(p)}{\rd p}\frac{\rd c_k}{\rd s}p^{-k}
        + \frac{\rd B_n(p)}{\rd s}kc_kp^{-k-1}\right),
\end{gather*}
which has f\/irst order poles at $p = b_i$,
an $N$-th order pole at $p = 0$ and no poles
in the f\/inite part of the Riemann sphere.
On the other hand, since $B^c_n(p) = O(p^{-1})$ as $p\to\infty$,
the complementary expression $\{\log\lambda(p),B^c_n(p)\}$
is also~$O(p^{-1})$.  Consequently, the right hand side
of the Lax equation is a rational function of the form
\begin{gather*}
  \{B_n(p),\log\lambda(p)\}
  = \{\log\lambda(p),B^c_n(p)\}
  = \sum_{i=1}^M\frac{\kappa_iF_{in}}{p-b_i}
    + \sum_{k=1}^N G_{kn}p^{-k},
\end{gather*}
where $F_{in}$'s and $G_{in}$'s are functions
of $b_i$'s, $c_k$'s and their f\/irst order derivatives
with respect to~$s$.  Thus the Lax equations
reduce to evolution equations as stated in the theorem.
\end{proof}

\subsection{Examples: two-variable reductions}\label{section2.3}

The simplest nontrivial cases are two-variable reductions.
They contain the following old and new examples
of integrable hierarchies.
\begin{itemize}\itemsep=0pt
\item[(i)] Landau--Ginzburg potential~(\ref{lambda(I)})
with $N = 1$, $M = 2$, $\kappa_1 = \kappa_2 = 1$:
\begin{gather*}
  \lambda(p) = p^{-1}(p-b_1)(p-b_2) = p + b + cp^{-1}.
\end{gather*}
This is the reduced Lax function of the dispersionless
1D Toda hierarchy.  It is well known that this hierarchy
plays a fundamental role in a wide range of issues
of mathematical physics.   In a literal sense,
this Laurent polynomial is used for the Landau--Ginzburg
(or ``mirror'') description of the topological
sigma model of~$\CC\PP^1$~\cite{EHY95, EY94}.
The associated Frobenius structure is also
a signif\/icant example of Dubrovin's duality \cite{Dubrovin04,RS07}.
\item[(ii)] Landau--Ginzburg potential (\ref{lambda(I)})
with $N = -1$, $M = 2$, $\kappa_1 = 1$, $\kappa_2 = -1$:
\begin{gather}
  \lambda(p) = p\frac{p - b}{p - c}.
\label{lambda-dAL}
\end{gather}
This is the reduced Lax function of the dispersionless
Ablowitz--Ladik hierarchy.  This hierarchy and
the dispersive version have found new applications
in universality classes of nonlinear waves \cite{Dubrovin08}
and local Gromov--Witten invariants of the resolved conifold
\cite{Brini10,BCR11}.
\item[(iii)] Landau--Ginzburg potential (\ref{lambda(II)})
with $N = M = 1$, $\kappa_1 = 1$:
\begin{gather*}
  \lambda(p) = (p - b)\exp\big(cp^{-1}\big).
\end{gather*}
This seems to give a new dispersionless integrable hierarchy.
\end{itemize}

As regards the third example, one can further impose
the condition $b = 0$ and obtain a~hierarchy
with the reduced Lax function
\begin{gather*}
  \lambda(p) = p\exp\big(cp^{-1}\big).
\end{gather*}
This is an interesting case in itself,
because the inverse function of $\lambda(p)$
is Lambert's $W$-function that plays a role
in the theory of Hurwitz numbers
(see our previous paper~\cite{Takasaki10}
and references cited therein).
This ``one-variable reduction'' can be generalized
to Landau--Ginzburg potentials of the form
\begin{gather}
  \lambda(p) = p^M\exp\left(\sum_{k=1}^Nc_kp^{-k}\right),
\label{lambda(III)}
\end{gather}
where $M$ is an arbitrary positive integer.
They should be classif\/ied as ``Case~III'',
though we shall not pursue this case in this paper.

\section{L\"owner equations}\label{section3}

\subsection{General scheme of f\/inite-variable reductions}\label{section3.1}

According to the general scheme~\cite{CLR09, Manas04,TT06,TTZ06},
f\/inite-variable reductions of the dispersionless Toda hierarchy
are characterized by the equations
\begin{gather}
  \frac{\rd z(p)}{\rd\lambda_n}
    = \frac{\alpha_n p}{p-\gamma_n}\frac{\rd z(p)}{\rd p},\qquad
  \frac{\rd\zbar(p)}{\rd\lambda_n}
    = \frac{\alpha_n p}{p-\gamma_n}\frac{\rd\zbar(p)}{\rd p}
\label{zzbar-Lowner-eq}
\end{gather}
(referred to as ``L\"owner equations'' in the following)
for the reduced Lax functions
\begin{gather}
  z(p) = z(p; \lambda_1,\ldots,\lambda_K),\qquad
  \zbar(p) = \zbar(p; \lambda_1,\ldots,\lambda_K).
\label{zzbar-ansatz}
\end{gather}
These equations are a variant of the radial L\"owner equations
that were f\/irst introduced by L\"owner~\cite{Lowner23}.

The reduced Lax functions depend on the space-time variables
through the reduced dynamical variables
$\lambda_n = \lambda_n(s,\bst,\bstbar)$.
The reduced dynamical variables, in turn, are required
to satisfy the ``hydrodynamic system''
\begin{gather}
  \frac{\rd\lambda_n}{\rd t_k}
    = V_{kn}\frac{\rd\lambda_n}{\rd s},\qquad
  \frac{\rd\lambda_n}{\rd\tbar_k}
    = \Vbar_{kn}\frac{\rd\lambda_n}{\rd s}.
\label{hd-system}
\end{gather}
The ``characteristic speeds'' $V_{kn} = V_{kn}(\lambda_1,\ldots,\lambda_K)$
and $\Vbar_{kn} = \Vbar_{kn}(\lambda_1,\ldots,\lambda_K)$
are def\/ined as
\begin{gather*}
  V_{kn} = \gamma_n B_k'(\gamma_n),\qquad
  \Vbar_{kn} = \gamma_n\Bbar_k'(\gamma_n),
\end{gather*}
where the prime denotes the derivative,
i.e., $B_k'(p) = \rd B_k(p)/\rd p$ and
$\Bbar_k'(p) = \rd\Bbar_k(p)/\rd p$.
Since $B_1(p) = p + u_1$, the characteristic speeds
for the $t_1$-f\/low coincides with $\gamma_n$:
\begin{gather*}
  V_{1n} = \gamma_n.
\end{gather*}

Thus, once the reduced Lax functions are given as a solution
of the L\"owner equations, the Lax equations are transformed
to the hydrodynamic system~(\ref{hd-system})
for the ``Riemann invariants'' $\lambda_1,\ldots,\lambda_K$.
A precise statement of this fact reads as follows~\cite{CLR09, Manas04,TT06,TTZ06}:

\begin{theorem}\label{thm:Lowner-eq}
If $z(p)$ and $\zbar(p)$ satisfy the L\"owner equations~\eqref{zzbar-Lownereq2} with respect to~$\lambda_n$'s,
and~$\lambda_n$'s satisfy the hydrodynamic system~\eqref{hd-system}
with respect to the space-time variables,
then~$z(p)$ and~$\zbar(p)$ satisfy the Lax equations~\eqref{dToda-Lax-eq} with respect to the space-time variables.
\end{theorem}

$\alpha_n$'s and $\gamma_n$'s in (\ref{zzbar-Lowner-eq})
are functions of $(\lambda_1,\dots,\lambda_K)$
to be determined in the reduction procedure.  Note that
they are not arbitrary functions.  Consistency
of (\ref{zzbar-Lowner-eq}) yield the dif\/ferential equations
\begin{gather}
  \frac{\rd\gamma_n}{\rd\lambda_m}
    = \frac{\alpha_m\gamma_n}{\gamma_m-\gamma_n}, \qquad
  \frac{\rd\alpha_n}{\rd\lambda_m}
    = \frac{\alpha_m\alpha_n(\gamma_m+\gamma_n)}
           {(\gamma_m-\gamma_n)^2}, \qquad m \not= n,
\label{GT-eq}
\end{gather}
which are referred as ``Gibbons--Tsarev equations''
or, more precisely, ``radial Gibbons--Tsarev equations''.
These equations are a radial version of
the celebrated Gibbons--Tsarev equa\-tions~\mbox{\cite{GT99, GT96}}
(see the remarks below).

It is interesting that the product and quotient
of $\alpha_n$, $\gamma_n$ satisfy the following equations
\begin{gather*}
  \frac{\rd}{\rd\lambda_m}(\alpha_n\gamma_n)
  = \frac{2(\alpha_m\gamma_m)(\alpha_n\gamma_n)}
    {(\gamma_m-\gamma_n)^2}, \qquad
  \frac{\rd}{\rd\lambda_m}\left(\frac{\alpha_n}{\gamma_n}\right)
  = \frac{2\gamma_m\gamma_n}{(\gamma_m-\gamma_n)^2}
    \frac{\alpha_m}{\gamma_m}\frac{\alpha_n}{\gamma_n}.%\label{GZ-eq2}
\end{gather*}
The right hand side of these equations are symmetric
with respect to~$m$ and~$n$.
This implies the existence of potentials.

It is easy to identify these potentials.
Expands both hand sides of~(\ref{zzbar-Lowner-eq})
into Laurent series at $p = \infty$ and pick out
the~$p^1$ and~$p^2$ terms from the f\/irst equation
and the~$p^0$ terms from the second equation.
One can thus f\/ind the relations
\begin{gather}
  \alpha_n = \frac{\rd u_1}{\rd\lambda_n},\qquad
  \alpha_n\gamma_n = \frac{\rd u_2}{\rd\lambda_n},\qquad
  \frac{\alpha_n}{\gamma_n} = \frac{\rd\phi}{\rd\lambda_n},
\label{potential}
\end{gather}
which show that $u_1$, $u_2$ and $\phi = \log\ubar_0$
play the role of potentials.
By the f\/irst expression of~$\alpha_n$ in~(\ref{potential}),
one can rewrite~(\ref{zzbar-Lowner-eq}) as
\begin{gather}
  \frac{\rd z(p)}{\rd\lambda_n}
  = \frac{pz'(p)}{p-\gamma_n}\frac{\rd u_1}{\rd\lambda_n},\qquad
  \frac{\rd\zbar(p)}{\rd\lambda_n}
  = \frac{p\zbar'(p)}{p-\gamma_n}\frac{\rd u_1}{\rd\lambda_n}.
\label{zzbar-Lownereq2}
\end{gather}
It is these equations that can be obtained directly
from the Lax equations under the f\/inite-variable ansatz~(\ref{zzbar-ansatz}).

The problem is now converted to solving (\ref{hd-system}).
This problem can be treated by the generalized hodograph method~\cite{Tsarev93, Tsarev85}.

\begin{remark}
The original form of the Gibbons--Tsarev equations
\cite{GT99, GT96} read
\begin{gather}
  \frac{\rd\gamma_n}{\rd\lambda_m}
    = \frac{\alpha_n}{\gamma_m-\gamma_n},\qquad
  \frac{\rd\alpha_n}{\rd\lambda_m}
    = \frac{2\alpha_m\alpha_n}{(\gamma_m-\gamma_n)^2}.
\label{chordal-GT-eq}
\end{gather}
They are integrability conditions of the chordal L\"owner equations
\begin{gather*}
  \frac{\rd z(p)}{\rd\lambda_n}
    = \frac{\alpha_n}{p-\gamma_n}\frac{\rd z(p)}{\rd p}
\end{gather*}
for the Lax function
\begin{gather*}
  z(p) = p + u_2p^{-1} + \cdots
\end{gather*}
of the Benney hierarchy or, more generally,
of the dispersionless KP hierarchy~\cite{MMAM02}.
Note that the $u_1$-term is absent here,
and $u_2$ plays the role of a potential
for the coef\/f\/icients $\alpha_n$
\begin{gather*}
  \alpha_n = \frac{\rd u_2}{\rd\lambda_n}.
\end{gather*}
\end{remark}

\begin{remark}
Ferapontov et al.~\cite{FKS02} presented
the radial Gibbons--Tsarev equations~(\ref{GT-eq})
(written in a trigonometric form) in their work
on f\/inite-variable reductions of the Boyer--Finley equation.
The Boyer--Finley equation is the lowest 2D part
of the dispersionless Toda hierarchy.
\end{remark}

\subsection{Hodograph solutions}\label{section3.2}

The generalized hodograph method \cite{Tsarev93, Tsarev85}
is based on the fact that the characteristic speeds
satisfy the equations
\begin{gather}
  \frac{1}{V_{km}-V_{kn}}\frac{\rd V_{kn}}{\rd\lambda_m}
  = \frac{1}{\Vbar_{km}-\Vbar_{kn}}\frac{\rd\Vbar_{kn}}{\rd\lambda_m}
  = \frac{\alpha_m\gamma_n}{(\gamma_m - \gamma_n)^2}
\label{hodograph-VVbar-eq}
\end{gather}
for $k = 1,2,\ldots$.  Note that these equations include
the special case
\begin{gather}
  \frac{1}{\gamma_m - \gamma_{n}}\frac{\rd\gamma_n}{\rd\lambda_m}
  = \frac{\alpha_m\gamma_n}{(\gamma_m - \gamma_n)^2}
\label{hodograph-gamma-eq}
\end{gather}
associated with $V_{1n} = \gamma_n$.
One can derive these equations directly from the def\/inition
of $B_k(p)$ and $\Bbar_k(p)$~\cite{Manas04}
(see the remarks below) or by generating functions
of these polyno\-mials~\mbox{\cite{CLR09, TT06}}.
Having these equations, one can readily apply
the generalized hodograph method
to the hydrodynamic system~(\ref{hd-system})
as follows~\cite{Manas04,TT06,TTZ06}:

\begin{theorem}
If a set of functions
$F_n = F_n(\lambda_1,\ldots,\lambda_K)$
satisfy the equations
\begin{gather}
  \frac{1}{F_m - F_n}\frac{\rd F_n}{\rd\lambda_m}
  = \frac{\alpha_m\gamma_n}{(\gamma_m - \gamma_n)^2}
\label{hodograph-F-eq}
\end{gather}
and the non-degeneracy condition
\begin{gather*}
  \det\left(\frac{\rd F_n}{\rd\lambda_m}\right)_{m,n=1,\dots, K} \not= 0,
\end{gather*}
then a solution of the hydrodynamic system \eqref{hd-system}
can be obtained from the hodograph relations
\begin{gather*}
  s + \sum_{k\ge 1}t_kV_{kn} + \sum_{k\ge 1}\tbar_k\Vbar_{kn}
  = F_n, \qquad n = 1,\ldots,K,
\end{gather*}
as a $K$-tuple of implicit functions $\lambda_n
= \lambda_n(s,\bst,\bstbar)$, $n = 1,\ldots,K$,
in a neighborhood of $(\bst,\bstbar) = (\bszero,\bszero)$.
\end{theorem}

\begin{remark}
The characteristic speeds $V_{kn}$ and $\Vbar_{kn}$
have the contour integral representation
\begin{gather*}
  V_{kn} = - \gamma_n\oint\frac{z(p)^n}{(p-\gamma_n)^2}
             \frac{dp}{2\pi i}, \qquad
  \Vbar_{kn} = - \gamma_n\oint\frac{\zbar(p)^n}{(p-\gamma_n)^2}
                 \frac{dp}{2\pi i},
\end{gather*}
where the contours encircle $p = \infty$ and $p = 0$, respectively,
leaving $\gamma_n$ outside.  One can use the L\"owner equations
and the Gibbons--Tsarev equations to dif\/ferentiate
these contour integrals.  Thus, after some algebra,
one can derive~(\ref{hodograph-VVbar-eq}).
\end{remark}

\begin{remark}
Solutions of~(\ref{hodograph-F-eq}), too, can be obtained
as contour integrals.  For example, arbitrary linear combinations
of $V_{kn}$ and $\Vbar_{kn}$, which are obvious solutions
of~(\ref{hodograph-F-eq}), can be cast into a~contour integral
of the form
\begin{gather*}
  F_n = \gamma_n\oint\frac{F_1(z(p))}{(p-\gamma_n)^2}\frac{dp}{2\pi i}
  + \gamma_n\oint\frac{F_2(\zbar(p))}{(p-\gamma_n)^2}\frac{dp}{2\pi i},
\end{gather*}
where $F_1(p)$ and $F_2(p)$ are arbitrary (analytic) functions.
If $z(p)$ and $\zbar(p)$ are obtained from a~globally def\/ined
Landau--Ginzburg potential $\lambda(p)$,
one can unify the two integrals to a single integral
\begin{gather*}
  F_n = \gamma_n\oint_{C}\frac{F(\lambda(p))}{(p-\gamma_n)^2}\frac{dp}{2\pi i}
\end{gather*}
along a general cycle $C$ in the domain of def\/inition of~$F(\lambda(p))$.
This gives a more general solution of~(\ref{hodograph-F-eq})
as Ferapontov et al.~\cite{FKS02} pointed out in their formulation.
\end{remark}

\subsection[L\"owner equations for Landau-Ginzburg potentials]{L\"owner equations for Landau--Ginzburg potentials}\label{section3.3}

If the Lax functions $z(p)$ and $\zbar(p)$ are reduced to
a single Landau--Ginzburg potential $\lambda(p)$,
the L\"owner equations (\ref{zzbar-Lowner-eq})
for the Lax functions, too, are reduced to the equations
\begin{gather}
  \frac{\rd\lambda(p)}{\rd\lambda_n}
  = \frac{\alpha_n p}{p - \gamma_n}\frac{\rd\lambda(p)}{\rd p}
\label{lambda-Lowner-eq}
\end{gather}
for $\lambda(p)$.

We show below that the Landau--Ginzburg potentials~(\ref{lambda(I)})
and~(\ref{lambda(II)}) do satisfy these equations.
The relevant variables $\lambda_n$ are the critical values
of $\lambda(p)$, i.e., the values of~$\lambda(p)$
at the critical points~$\gamma_n$'s,
\begin{gather}\label{add1}
  \lambda_n = \lambda(\gamma_n),\qquad
  \lambda'(\gamma_n) = 0, \qquad n = 1,\ldots,K,
\end{gather}
where $K = M$ in the case of~(\ref{lambda(I)})
and $K = M+N$ in the case of~(\ref{lambda(II)}).
We choose these $\lambda_n$'s as new coordinates
on the parameter space of $\lambda(p)$, and treat~$\lambda(p)$
as a function $\lambda(p;\lambda_1,\ldots,\lambda_K)$
of~$p$ and~$\lambda_n$'s.

Let us show a few technical remarks.

\begin{lemma}
\begin{gather}
    \left.\frac{\rd\lambda(p)}{\rd\lambda_m}\right|_{p=\gamma_n}
    = \delta_{mn}.
\label{d-lambda(gamma)}
\end{gather}
\end{lemma}

\begin{proof}
By the def\/inition \eqref{add1} of $\lambda_n$'s and the chain rule of dif\/ferentiation,
\begin{gather*}
  \delta_{mn} = \frac{\rd\lambda_n}{\rd\lambda_m}
  = \left.\frac{\rd\lambda(p)}{\rd\lambda_m}\right|_{p=\gamma_n}
    + \lambda'(\gamma_m)
  = \left.\frac{\rd\lambda(p)}{\rd\lambda_m}\right|_{p=\gamma_n}. \tag*{\qed}
\end{gather*}
\renewcommand{\qed}{}
\end{proof}

\begin{lemma}
If the L\"owner equations \eqref{lambda-Lowner-eq} are satisfied,
the coefficients $\alpha_n$ are uniquely determined
by the equations themselves as
\begin{gather}
  \alpha_n = \frac{1}{\gamma_n\lambda''(\gamma_n)}.
\label{alpha-formula}
\end{gather}
\end{lemma}

\begin{proof}
Let $p \to \gamma_n$ in (\ref{lambda-Lowner-eq}).
By (\ref{d-lambda(gamma)}), the left hand side
tends to $1$.  As regards the right hand side,
\begin{gather*}
  \lim_{p\to\gamma_n}\frac{\alpha_np}{p-\gamma_n}\frac{\rd\lambda(p)}{\rd p}
  = \alpha_n\gamma_n\lim_{p\to\gamma_n}\frac{\lambda'(p)}{p-\gamma_n}
  = \alpha_n\gamma_n\lambda''(\gamma_n). \tag*{\qed}
\end{gather*}
  \renewcommand{\qed}{}
\end{proof}

Bearing these technical remarks in mind, let us examine
the two cases separately.

{\bf Case~I.}
It is convenient to consider the logarithmic derivative,
rather than the derivative,
of the Landau--Ginzburg potential (\ref{lambda(I)}):
\begin{gather}
  \frac{\rd\log\lambda(p)}{\rd p}
  = - \frac{N}{p} + \sum_{i=1}^M\frac{\kappa_i}{p-b_i}
  = \frac{Q(p)}{p\prod_{i=1}^M(p-b_i)}.
\label{dlog(p)-lambda(I)}
\end{gather}
The numerator $Q(p)$ is a polynomial of the form
\begin{gather*}
  Q(p) = \Mtilde p^M + \cdots.
\end{gather*}
We assume that $Q(p)$ has $M$ distinct zeroes
$\gamma_n$, $n = 1,\ldots,M$,
\begin{gather*}
  Q(p) = \Mtilde\prod_{n=1}^M(p - \gamma_n), \qquad
  \gamma_m \not= \gamma_n \qquad \text{for} \quad m \not= n.
\end{gather*}
From now on, $\lambda(p)$ is understood to be a function
of~$p$ and~$\lambda_n$'s.  The parameters $b_i$'s, too,
become functions of~$\lambda_n$'s.
The derivative of $\log\lambda(p)$ with respect to $\lambda_m$
can be expressed as
\begin{gather}
  \frac{\rd\log\lambda(p)}{\rd\lambda_m}
  = - \sum_{i=1}^M\frac{\kappa_i}{p-b_i}\frac{\rd b_i}{\rd\lambda_m}
  = \frac{Q_m(p)}{\prod_{i=1}^M(p-b_i)},
\label{dlog(lam)-lambda(I)}
\end{gather}
where $Q_m(p)$ is a polynomial of degree less than~$M$.
(\ref{dlog(p)-lambda(I)})~and~(\ref{dlog(lam)-lambda(I)})
imply the equality
\begin{gather*}
  \frac{\rd\log\lambda(p)}{\rd\lambda_m}
  = \frac{Q_m(p)p}{Q(p)}\frac{\rd\log\lambda(p)}{\rd p}.
\end{gather*}
The problem is to f\/ind an explicit form of the pre-factor $Q_m(p)p/Q(p)$.

Since (\ref{d-lambda(gamma)}) implies that
\begin{gather*}
  \left.\frac{\rd\log\lambda(p)}{\rd\lambda_m}\right|_{p=\gamma_n}
  = \left.\frac{1}{\lambda(p)}
    \frac{\rd\lambda(p)}{\rd\lambda_m}\right|_{p=\gamma_n}
  = \frac{\delta_{mn}}{\lambda_n},
\end{gather*}
letting $p \to \gamma_n$ in (\ref{dlog(lam)-lambda(I)})
yields that
\[
  Q_m(\gamma_n) = 0 \qquad\text{for}\quad   n \not= m .
\]
Therefore, by the Lagrange interpolation formula,
$Q_m(p)/Q(p)$ can be expressed as
\begin{gather*}
  \frac{Q_m(p)}{Q(p)}
  = \sum_{n=1}^M\frac{Q_m(\gamma_n)}{Q'(\gamma_n)(p-\gamma_n)}
  = \frac{Q_m(\gamma_m)}{Q'(\gamma_n)(p-\gamma_m)}.
\end{gather*}
Thus, def\/ining $\alpha_m$ as
\begin{gather*}
  \alpha_m = Q_m(\gamma_m)/Q'(\gamma_m),
\end{gather*}
we obtain the L\"owner equations
\begin{gather*}
  \frac{\rd\log\lambda(p)}{\rd\lambda_m}
  = \frac{\alpha_m p}{p-\gamma_m}\frac{\rd\log\lambda(p)}{\rd p}
\end{gather*}
for $\log\lambda(p)$.  Of course, they are equivalent
to the L\"owner equations~(\ref{lambda-Lowner-eq}) for $\lambda(p)$.
By the second lemma above, $\alpha_m$ turns out
to have another expression~(\ref{alpha-formula}).

{\bf Case~II.}
The logarithmic derivatives of the Landau--Ginzburg potential
(\ref{lambda(II)}) can be expresses as
\begin{gather}
  \frac{\rd\log\lambda(p)}{\rd p}
  = \sum_{i=1}^M\frac{\kappa_i}{p-b_i} - \sum_{k=1}^N kc_kp^{-k-1}
  = \frac{Q(p)}{p^{N+1}\prod_{i=1}^M(p-b_i)}
\label{dlog(p)-lambda(II)}
\end{gather}
and
\begin{gather*}
  \frac{\rd\log\lambda(p)}{\rd\lambda_m}
  = - \sum_{i=1}^M\frac{\kappa_i}{p-b_i}\frac{\rd b_i}{\rd\lambda_m}
    + \sum_{k=1}^N\frac{\rd c_k}{\rd\lambda_m}p^{-k}
  = \frac{Q_m(p)}{p^N\prod_{i=1}^M(p-b_i)},
\end{gather*}
where $Q(p)$ is a polynomial of the form
\begin{gather*}
  Q(p) = \Mtilde\prod_{n=1}^{M+N}(p - \gamma_n), \qquad
 \gamma_m \not= \gamma_n \qquad \text{for}\quad m \not= n .
\end{gather*}
$Q_m(p)$ is a polynomial of degree less than $M+N$,
and the roots $\gamma_n$ of $Q(p)$ are assumed to be distinct.
Starting from these data, one can derive
the L\"owner equations~(\ref{lambda-Lowner-eq})
in much the same way as in the Case~I.

We thus obtain the following result:

\begin{theorem}
$\lambda(p) = \lambda(p;\lambda_1,\ldots,\lambda_K)$
satisfies the L\"owner equations \eqref{lambda-Lowner-eq}
with the coefficients~$\alpha_n$ defined by~\eqref{alpha-formula}.
\end{theorem}

On the basis of this result, we can apply the foregoing scheme
of f\/inite-variable reductions to the Landau--Ginzburg potentials~(\ref{lambda(I)}) and~(\ref{lambda(II)}).

\section{Darboux equations}\label{section4}

\subsection{Basic notions in classical dif\/ferential geometry}\label{section4.1}

Given a diagonal metric $ds^2 = \sum\limits_{n=1}^K (h_nd\lambda_n)^2$,\footnote{We do not use Einstein's convention in this paper.}
one can def\/ine the rotation coef\/f\/icients $\beta_{mn}$, $m \not= n$, as
\begin{gather*}
  \beta_{mn} = \frac{1}{h_m}\frac{\rd h_n}{\rd\lambda_m}.
\end{gather*}
$h_n$'s are called ``Lam\'e coef\/f\/icients'' in the theory  of
orthogonal curvilinear coordinate sys\-tems~\mbox{\cite{Darboux-oclc,Tsarev93}}.
The Riemann curvature of this metric vanishes
if and only if the following equations are satisf\/ied
\begin{gather}
  \frac{\rd\beta_{mn}}{\rd\lambda_k} = \beta_{mk}\beta_{kn}\qquad
  \text{for}\quad  k \not= m,n,
  \label{Darboux-eq}\\
  \frac{\rd\beta_{mn}}{\rd\lambda_m}
  + \frac{\rd\beta_{mn}}{\rd\lambda_n}
  + \sum_{k=1}^K\beta_{km}\beta_{kn} = 0.
  \label{flatness-eq}
\end{gather}
The f\/irst part (\ref{Darboux-eq}) of these equations
are called ``Darboux equations'' in the literature.
Thus the Darboux equations are partial-f\/latness conditions,
and have to be supplemented by the second equations~(\ref{flatness-eq})
to ensure f\/latness.

If the rotation coef\/f\/icients are symmetric,
i.e., $\beta_{mn} = \beta_{nm}$,
the metric components satisfy the conditions
(Egorov conditions)
\begin{gather*}
  \frac{\rd}{\rd\lambda_m}\big(h_n^2\big)
  = \frac{\rd}{\rd\lambda_n}\big(h_m^2\big)
\end{gather*}
that ensure the existence of a potential $\phi$
(Egorov potential) such that
\begin{gather*}
  h_n^2 = \frac{\rd\phi}{\rd\lambda_n}.
\end{gather*}
If the Darboux equations and the Egorov condition
are satisf\/ied, the f\/latness condition~(\ref{flatness-eq})
reduces to the equations
\begin{gather}
  \sum_{k=1}^K\frac{\rd\beta_{mn}}{\rd\lambda_k} = 0.
\label{flatness-eq2}
\end{gather}
A diagonal metric that satisf\/ies the Darboux equations
and the Egorov conditions is called an ``Egorov metric''.
Thus an Egorov metric is associated with a symmetric
($\beta_{mn} = \beta_{nm}$) solution of the coupled system
of the Darboux equations and~(\ref{flatness-eq2}).
This system is closely related to the $K$-wave system~\cite{Dubrovin-2dtft, Dubrovin90}.

If one starts from a solution of the Darboux equations
(\ref{Darboux-eq}), the Lam\'e coef\/f\/icients
are reco\-ve\-red as a solution of the equations
\begin{gather*}
  \frac{\rd h_n}{\rd\lambda_m} = h_m\beta_{mn}.
%\label{Lame-eq}
\end{gather*}
The Darboux equations are integrability conditions
of these linear equations.  These equations leaves
some arbitrariness in the Lam\'e coef\/f\/icients.
Two sets $h_n$, $\htilde_n$ of Lam\'e coef\/f\/icients
have the same rotation coef\/f\/icients if and only if
their ratios $w_n = \htilde_n/h_n$ satisfy the equations
\begin{gather}
  \frac{1}{w_m-w_n}\frac{\rd w_n}{\rd\lambda_m}
  = \frac{\rd\log h_n}{\rd\lambda_m}.
\label{Combescure-eq}
\end{gather}
Any solution of these equations thus gives a transformation
on the set of Lam\'e coef\/f\/icients
with the same rotational coef\/f\/icients.
This transformation is called ``Combescure transformation''
in the theory  of orthogonal curvilinear coordinate systems~\cite{Darboux-oclc,Tsarev93}.

\subsection[Implications of Gibbons-Tsarev equations]{Implications of Gibbons--Tsarev equations}\label{section4.2}

Let us consider the Gibbons--Tsarev equations (\ref{GT-eq})
in the language of Darboux equations and Egorov metrics.
There are three metrics that are of particular interest:
\begin{gather}
  \sum_{n=1}^K (h_nd\lambda_n)^2,\qquad
  h_n = \sqrt{\alpha_n/\gamma_n},
\label{metric(h)}\\
  \sum_{n=1}^K (\htilde_nd\lambda_n)^2,\qquad
  \htilde_n = \sqrt{\alpha_n\gamma_n},
\label{metric(htilde)}\\
  \sum_{n=1}^K (\hhat_nd\log\lambda_n)^2,\qquad
  \hhat_n = \sqrt{\alpha_n\lambda_n/\gamma_n}.
\label{metric(hhat)}
\end{gather}
(\ref{metric(h)}) and (\ref{metric(htilde)}) underlie
the hodograph solutions of the hydrodynamic equations (\ref{hd-system}).
(\ref{metric(h)}) and (\ref{metric(hhat)}) are the metrics
that we shall consider in the next section
in the context of Frobenius structures.

We show below that the rotation coef\/f\/icients of these metrics
are symmetric and satisfy the Darboux equations
(with respect to~$\lambda_n$'s in the f\/irst and second
cases and $\log\lambda_n$'s in the third case).
Egorov potentials themselves can be readily identif\/ied
as one can see from~(\ref{potential}):
\begin{gather*}
  h_n{}^2 = \frac{\rd\phi}{\rd\lambda_n},\qquad
  \htilde_n{}^2 = \frac{\rd u_2}{\rd\lambda_n},\qquad
  \hhat_n{}^2 = \frac{\rd\phi}{\rd\log\lambda_n}.
\end{gather*}

The f\/irst two cases (\ref{metric(h)}) and (\ref{metric(htilde)})
are closely related.

\begin{theorem}
The Lam\'e coefficients of \eqref{metric(h)} and \eqref{metric(htilde)}
have the same rotation coefficients
\begin{gather}
  \beta_{mn} = \beta_{nm}
  = \frac{\sqrt{\alpha_m\alpha_n\gamma_m\gamma_n}}
         {(\gamma_m-\gamma_n)^2}.
\label{beta(h)}
\end{gather}
These rotation coefficients satisfy the Darboux equations~\eqref{Darboux-eq}.
\end{theorem}

\begin{proof}
Do straightforward calculations using the Gibbons--Tsarev equations
(\ref{GT-eq}).  The rotation coef\/f\/icients of $h_n$'s
can be calculated as
\begin{gather*}
  \frac{1}{h_m}\frac{\rd h_n}{\rd\lambda_m}
  = \frac{h_n}{h_m}\frac{\rd\log h_n}{\rd\lambda_m}
   = \sqrt{\frac{\alpha_n\gamma_m}{\alpha_m\gamma_n}}
     \left(\frac{1}{2\alpha_n}\frac{\rd\alpha_n}{\rd\lambda_m}
         - \frac{1}{2\gamma_n}\frac{\rd\gamma_n}{\rd\lambda_m}
     \right)\\
\hphantom{\frac{1}{h_m}\frac{\rd h_n}{\rd\lambda_m}}{}
= \sqrt{\frac{\alpha_n\gamma_m}{\alpha_m\gamma_n}}
     \left(\frac{1}{2\alpha_n}
           \frac{\alpha_m\alpha_n(\gamma_m+\gamma_n)}
                {(\gamma_m-\gamma_n)^2}
         - \frac{1}{2\gamma_n}
           \frac{\alpha_m\gamma_n}{\gamma_m-\gamma_n}
     \right)
 = \frac{\sqrt{\alpha_m\alpha_n\gamma_m\gamma_n}}
          {(\gamma_m-\gamma_n)^2}.
\end{gather*}
In much the same way, the rotation coef\/f\/icients
of $\htilde_n$'s can be calculated as
\begin{gather*}
  \frac{1}{\htilde_n}\frac{\rd\htilde_n}{\rd\lambda_m}
   = \frac{\htilde_n}{\htilde_m}\frac{\rd\log\htilde_n}{\rd\lambda_m}
   = \sqrt{\frac{\alpha_n\gamma_n}{\alpha_m\gamma_m}}
     \left(\frac{1}{2\alpha_n}\frac{\rd\alpha_n}{\rd\lambda_m}
         + \frac{1}{2\gamma_n}\frac{\rd\gamma_n}{\rd\lambda_m}
     \right)\\
\hphantom{\frac{1}{\htilde_n}\frac{\rd\htilde_n}{\rd\lambda_m}}{}
= \sqrt{\frac{\alpha_n\gamma_n}{\alpha_m\gamma_m}}
     \left(\frac{1}{2\alpha_n}
           \frac{\alpha_m\alpha_n(\gamma_m+\gamma_n)}
                {(\gamma_m-\gamma_n)^2}
         + \frac{1}{2\gamma_n}
           \frac{\alpha_m\gamma_n}{\gamma_m-\gamma_n}
     \right)
 = \frac{\sqrt{\alpha_m\alpha_n\gamma_m\gamma_n}}
          {(\gamma_m-\gamma_n)^2}.
\end{gather*}
Thus the rotation coef\/f\/icients of $h_n$'s and $\htilde_n$'s
turn out to coincide.  Dif\/ferentiating them with respect
to~$\lambda_k$ and doing some algebra, one can derive
the Darboux equations.
\end{proof}

\begin{corollary}
$\htilde_n$'s are a Combescure transformation
of~$h_n$'s, and the ratios $\gamma_n = \htilde_n/h_n$
sa\-tis\-fy~\eqref{Combescure-eq}.
\end{corollary}

The right hand side of (\ref{Combescure-eq}) can be
calculated explicitly as
\begin{gather*}
  \frac{\rd\log h_n}{\rd\lambda_m}
  = \frac{\alpha_m\gamma_n}{(\gamma_m-\gamma_n)^2},
\end{gather*}
thus (\ref{Combescure-eq}) coincides with
(\ref{hodograph-gamma-eq}).  Since
(\ref{hodograph-gamma-eq}) is a special case of
(\ref{hodograph-VVbar-eq}), the characteristic speeds~$V_{kn}$ and~$\Vbar_{kn}$, too, generate
Combescure transformations of~$h_n$.  An analogous fact
is known for the dispersionless KP hierarchy~\cite{MMAM02} and universal Whitham hierarchy~\cite{GMMA03} as well (see the remark below).
Thus, as stressed by Tsarev~\cite{Tsarev93},
the notion of Combescure transformations lies
in the heart of integrability of
hydrodynamic systems such as~(\ref{hd-system}).

The third case (\ref{metric(hhat)}) is of somewhat
dif\/ferent nature.  It is $\log\lambda_n$'s
rather than~$\lambda_n$'s that are used
to formulate the Darboux equations.

\begin{theorem}
The rotation coefficients
\begin{gather*}
    \betahat_{mn}
    = \frac{1}{\hhat_m}\frac{\rd\hhat_n}{\rd\log\lambda_m},\qquad
    m \not= n,
\end{gather*}
of the Lam\'e coefficients of \eqref{metric(hhat)}
are related to \eqref{beta(h)} as
\begin{gather*}
  \betahat_{mn} = \sqrt{\lambda_m\lambda_n}\beta_{mn},
\end{gather*}
and satisfy the Darboux equations
\begin{gather}
  \frac{\rd\betahat_{mn}}{\rd\log\lambda_k}
  + \betahat_{mk}\betahat_{kn}
  = 0 \qquad \text{for}\quad k \not= m,n.
\label{log-Darboux-eq}
\end{gather}
\end{theorem}

\begin{proof}
Do straightforward calculations.  Note, in particular,
that the left hand side of~(\ref{log-Darboux-eq})
and~(\ref{Darboux-eq}) are related as
\begin{gather*}
  \frac{\rd\betahat_{mn}}{\rd\log\lambda_k}
    + \betahat_{mk}\betahat_{kn}
  = \sqrt{\lambda_m\lambda_n}\lambda_k
    \left(\frac{\rd\beta_{mn}}{\rd\lambda_k}
      + \beta_{mk}\beta_{kn}\right).\tag*{\qed}
\end{gather*}
  \renewcommand{\qed}{}
\end{proof}

\begin{remark}
\label{rem:hhat-flatness}
If the rotation coef\/f\/icients $\betahat_{mn}$
are homogeneous functions of $\lambda_k$'s
of degree zero, the Darboux equations
(\ref{log-Darboux-eq}) and the Euler equations
\begin{gather*}
  \sum_{k=1}^N\frac{\rd\betahat_{mn}}{\rd\log\lambda_k}
  = \sum_{k=1}^N\lambda_k\frac{\rd\betahat_{mn}}{\rd\lambda_k}
  = 0
\end{gather*}
imply f\/latness of (\ref{metric(hhat)}).
This is indeed the case for the metrics $(\;,\;)$
considered in the next section.
\end{remark}

\begin{remark}
In the chordal case (\ref{chordal-GT-eq}),
the two sets
$
  h_n = \sqrt{\alpha_n}$, $
  \htilde_n = \sqrt{\alpha_n}\gamma_n$
of Lam\'e coef\/f\/icients amount to (\ref{metric(h)})
and (\ref{metric(htilde)}).
They have the same rotation coef\/f\/icients
\begin{gather*}
  \beta_{mn} = \beta_{nm}
  = \frac{\sqrt{\alpha_m\alpha_n}}{(\gamma_m-\gamma_n)^2}
\end{gather*}
that satisfy the Darboux equations (\ref{Darboux-eq}).
Consequently, $\htilde_n$'s are a Combescure transformation
of~$h_n$'s, and the ratios $\htilde_n/h_n = \gamma_n$
satisfy the equations
\begin{gather*}
  \frac{1}{\gamma_m-\gamma_n}\frac{\rd\gamma_n}{\rd\lambda_m}
  = \frac{\rd\log h_n}{\rd\lambda_m}
  = \frac{\alpha_m}{(\gamma_m-\gamma_n)^2}.
\end{gather*}
The last equations are fundamental equations
in the hodograph solutions of the Benney equations~\cite{GT99, GT96},
the dispersionless KP hierarchy~\cite{MMAM02}
and the universal Whitham hierarchy~\cite{GMMA03,TT08}.
\end{remark}

\section{Flat coordinates}\label{section5}

\subsection{Flat coordinates in Case~I}\label{section5.1}

Let us recall Dubrovin and Zhang's construction~\cite{DZ98}
of two Frobenius structures on the parameter space
of the Landau--Ginzburg potential
\begin{gather}
  \lambda(p) = p^{-N}\prod_{i=1}^N(p - b_i).
\label{DZ-lambda}
\end{gather}
The Frobenius structures are realized by
the following inner products (or metrics)
$\langle\;,\;\rangle$, $(\;,\;)$ and cubic forms
$\langle\;,\;,\;\rangle$, $(\;,\;,\;)$ for vector f\/ields
on the parameter space of the Laurent polynomial:
\begin{gather}
  \langle\rd,\rd'\rangle
  = \sum_{n=1}^M \res_{p=\gamma_n}\left[
      \frac{\rd\lambda(p)\cdot\rd'\lambda(p)}{d\lambda(p)}
      (d\log p)^2\right], \label{<,>(I)}\\
  \langle\rd,\rd',\rd''\rangle
  = \sum_{n=1}^M \res_{p=\gamma_n}\left[
      \frac{\rd\lambda(p)\cdot\rd'\lambda(p)
            \cdot\rd''\lambda(p)}{d\lambda(p)}
      (d\log p)^2\right], \label{<,,>(I)}\\
  (\rd,\rd')
  = \sum_{n=1}^M \res_{p=\gamma_n}\left[
      \frac{\rd\log\lambda(p)\cdot\rd'\log\lambda(p)}{d\log\lambda(p)}
      (d\log p)^2\right], \label{(,)(I)}\\
  (\rd,\rd',\rd'')
  = \sum_{n=1}^M \res_{p=\gamma_n}\left[
      \frac{\rd\log\lambda(p)\cdot\rd'\log\lambda(p)
            \cdot\rd''\log\lambda(p)}{d\log\lambda(p)}
      (d\log p)^2\right]. \label{(,,)(I)}
\end{gather}
The cubic forms are used to def\/ine two commutative and
associative product structures~$\circ$,~$\star$ of vector f\/ields
\begin{gather*}
  \langle\rd\circ\rd',\rd''\rangle
  = \langle\rd,\rd'\circ\rd''\rangle
  = \langle\rd,\rd',\rd''\rangle, \qquad
  (\rd\star\rd',\rd'')
  = (\rd,\rd'\star\rd'')
  = (\rd,\rd',\rd'').
\end{gather*}
These two Frobenius structures are a prototype
of Dubrovin's duality~\cite{Dubrovin04,RS07}.

When $\rd$, $\rd'$, $\rd''$ are derivatives in $\lambda_n$'s,
one can use the L\"owner equations (\ref{lambda-Lowner-eq})
to evaluate these inner products and cubic forms as follows
\begin{gather}
  \left\langle\frac{\rd}{\rd\lambda_m},
              \frac{\rd}{\rd\lambda_n}\right\rangle
  = \delta_{mn}\frac{\alpha_n}{\gamma_n},\qquad
  \left\langle\frac{\rd}{\rd\lambda_k},
              \frac{\rd}{\rd\lambda_m},
              \frac{\rd}{\rd\lambda_n}\right\rangle
  = \delta_{kmn}\frac{\alpha_n}{\gamma_n},
  \label{<,>-lambda}\\
  \left(\frac{\rd}{\rd\lambda_m},
        \frac{\rd}{\rd\lambda_n}\right)
  = \delta_{mn}\frac{\alpha_n}{\gamma_n\lambda_n},\qquad
  \left(\frac{\rd}{\rd\lambda_k},
        \frac{\rd}{\rd\lambda_m},
        \frac{\rd}{\rd\lambda_n}\right)
  = \delta_{kmn}\frac{\alpha_n}{\gamma_n\lambda_n{}^2},
  \label{(,)-lambda}
\end{gather}
where $\delta_{kmn} = \delta_{km}\delta_{mn}$
(i.e., $\delta_{kmn}$ is equal to $1$ if $k=m=n$
and $0$ otherwise).  Thus~(\ref{<,>(I)}) and~(\ref{(,)(I)})
correspond to the Egorov metrics~(\ref{metric(h)}) and~(\ref{metric(hhat)})
considered in the last section.
Note that~(\ref{<,>-lambda}) and~(\ref{(,)-lambda})
hold as far as the L\"owner equations are satisf\/ied.
In particular, they are valid for the general case
of~(\ref{lambda(I)}) as well.

As shown by Dubrovin and Zhang~\cite{DZ98},
the f\/irst metric~(\ref{<,>(I)}) has a system
of f\/lat coordinates $q_1,\ldots,q_{M-1}$,
$\qbar_0,\qbar_1,\ldots,\qbar_N$ def\/ined by
the residue formula
\begin{gather*}
  q_n = \res_{p=\infty}\left[\frac{z(p)^n}{n}d\log p\right],\qquad
  \qbar_0 = \phi,\qquad
  \qbar_n = \res_{p=0}\left[\frac{\zbar(p)^n}{n}d\log p\right].
\end{gather*}
The inner products of $\rd/\rd q_n$'s and $\rd/\rd\qbar_n$'s
are calculated explicitly as
\begin{gather*}
  \left\langle\frac{\rd}{\rd q_m},\frac{\rd}{\rd q_n}\right\rangle
  = \Mtilde\delta_{m+n,\Mtilde}, \qquad
  \left\langle\frac{\rd}{\rd\qbar_m},\frac{\rd}{\rd\qbar_n}\right\rangle
  = N\delta_{m+n,N},\qquad
  \left\langle\frac{\rd}{\rd q_m},\frac{\rd}{\rd\qbar_n}\right\rangle
  = 0.
\end{gather*}

Unfortunately, this construction of f\/lat coordinates
does not work for the more general Landau--Ginzburg potential~(\ref{lambda(I)}).  Actually, it seems likely that
the metric (\ref{<,>(I)}) is no longer f\/lat in other cases\footnote{We thank one of the referees
for pointing out this possibility.}.
In this respect, it is very signif\/icant that
Brini et al.~\cite{BCR11} extended Dubrovin and Zhang's
dual pair of Frobenius structures to the Lax function~(\ref{lambda-dAL}) of the dispersionless Ablowitz--Ladik hierarchy.

On the other hand, the construction of
the second Frobenius structure is valid
for the general case of~(\ref{lambda(I)}) as well.
This fact is shown by Chang~\cite{Chang07dtoda}
in the case where $N = -1$ and $\kappa_i$'s are arbitrary.
$\log b_i$'s are f\/lat coordinates of
the second metric~(\ref{(,)(I)}).
One can conf\/irm, with slightest modif\/ication
of Chang's calculations, that this is also the case
for an arbitrary value of $N$ as the following result
of calculations of the inner product shows
\begin{gather*}
  \left(\frac{\rd}{\rd b_i},\frac{\rd}{\rd b_j}\right)
  = (1 - \delta_{ij})\frac{\kappa_i\kappa_j}{Nb_ib_j}
    + \delta_{ij}\frac{(\kappa_i-N)\kappa_i}{Nb_i{}^2}.
\end{gather*}
We omit details of these calculations, which are parallel
to the proof of Lemma 3 below.  Let us note
that f\/latness of~(\ref{(,)(II)}) is also a consequence
of homogeneity of the rotation coef\/f\/icients
(cf.\ Remark~\ref{rem:hhat-flatness}).

\begin{remark}
Speaking more precisely, the def\/inition of a Frobenius manifold
requires some more data, in particular, an Euler vector f\/ield~$E$
and associated scaling properties~\cite{Dubrovin-2dtft}.
In the present setting, the Landau--Ginzburg potential~$\lambda(p)$
has natural homogeneity, and one can use the vector f\/ield
\begin{gather*}
  E = \sum_{n=1}^M \lambda_n\frac{\rd}{\rd\lambda_n}
    = \frac{1}{\Mtilde}\sum_{i=1}^M b_i\frac{\rd}{\rd b_i}
\end{gather*}
as an Euler vector f\/ield.  Furthermore, one can introduce
the prepotential $\mathcal{F}$ as a function of
the f\/lat coordinates $t_n$ and express the cubic form as
\begin{gather*}
  \left(\frac{\rd}{\rd t_l},\frac{\rd}{\rd t_m},\frac{\rd}{\rd t_n}
  \right) = \frac{\rd^3\mathcal{F}}{\rd t_l\rd t_m\rd t_n}.
\end{gather*}
In the following, we refer to the notion of Frobenius manifolds
in a loose sense, and focus our consideration on f\/latness of metrics.
\end{remark}

\subsection{Flat coordinates in Case~II}\label{section5.2}

We now turn to the Landau--Ginzburg potential~(\ref{lambda(II)}).  The goal is to present a set
of f\/lat coordinates for the inner product
\begin{gather}
  (\rd,\rd')
  = \sum_{n=1}^{M+N} \res_{p=\gamma_n}\left[
      \frac{\rd\log\lambda(p)\cdot\rd'\log\lambda(p)}{d\log\lambda(p)}
      (d\log p)^2\right].
\label{(,)(II)}
\end{gather}
This inner product corresponds to the Egorov metric~(\ref{metric(hhat)}).  Its f\/latness is ensured
by homogeneity of the rotation coef\/f\/icients
(cf.\ Remarks~\ref{rem:hhat-flatness} and~\ref{rem:II-homogeneity}).
The associated cubic form is def\/ined by
\begin{gather*}
  (\rd,\rd',\rd'')
  = \sum_{n=1}^{M+N} \res_{p=\gamma_n}\left[
      \frac{\rd\log\lambda(p)\cdot\rd'\log\lambda(p)
            \cdot\rd''\log\lambda(p)}{d\log\lambda(p)}
      (d\log p)^2\right],
%\label{(,,)(II)}
\end{gather*}
though we shall not study its implications.
Let us mention that the technical details and the f\/inal result
of the following consideration are remarkably similar
to the case of Ferguson and Strachan~\cite{FS06}.

Let us start from the natural coordinates
$b_1,\ldots,b_M$, $c_1,\ldots,c_N$ of the parameter space.
One can calculate part of the inner product explicitly
as follows.

\begin{lemma}
\begin{gather*}
  \left(\frac{\rd}{\rd b_i},\frac{\rd}{\rd b_j}\right)
  = - \delta_{ij}\frac{\kappa_i}{b_i{}^2},\qquad
  \left(\frac{\rd}{\rd b_i},\frac{\rd}{\rd c_k}\right)
  = \delta_{kN}\frac{\kappa_i}{Nb_i c_N}.
\end{gather*}
\end{lemma}

\begin{proof}
The derivative of $\log\lambda(p)$ with respect to $p$
is a function as shown in (\ref{dlog(p)-lambda(II)}).
The derivatives with respect to $b_i$ and $c_k$
take the simple form
\begin{gather*}
  \frac{\rd\log\lambda(p)}{\rd b_i} = - \frac{\kappa_i}{p-b_i},\qquad
  \frac{\rd\log\lambda(p)}{\rd c_k} = p^{-k}.
\end{gather*}
Consequently, the inner products in question can be expressed as
\begin{gather*}
  \left(\frac{\rd}{\rd b_i},\frac{\rd}{\rd b_j}\right)
  = \sum_{n=1}^{M+N}\res_{p=\gamma_n}
      \left[\frac{\kappa_i \kappa_j}{(p-b_i)(p-b_j)}
        \left(\frac{\rd\log\lambda(p)}{\rd p}\right)^{-1}
      \frac{dp}{p^2}\right],\\
  \left(\frac{\rd}{\rd b_i},\frac{\rd}{\rd c_k}\right)
  = \sum_{n=1}^{M+N}\res_{p=\gamma_n}
      \left[-\frac{\kappa_ip^{-k}}{p-b_i}
        \left(\frac{\rd\log\lambda(p)}{\rd p}\right)^{-1}
      \frac{dp}{p^2}\right].
\end{gather*}
Since
\begin{gather*}
  \left(\frac{\rd\log\lambda(p)}{\rd p}\right)^{-1}
  = \frac{p^{N+1}\prod\limits_{k=1}^M(p-b_k)}{Q(p)}, \qquad
  Q(p) = \Mtilde\prod_{n=1}^{M+N}(p-\gamma_n),
\end{gather*}
the 1-forms in the residues are rational
and have poles of the f\/irst order at $p = \gamma_1,\ldots,\gamma_{M+N}$.
Other possible poles are located at $p = b_i,b_j,0$.
The latter poles, however, can disappear because of
zeros of the numerator in this expression
of $(\rd\log\lambda(p)/\rd p)^{-1}$.
For example, if $i \not= j$, the f\/irst 1-form
is non-singular at $p = b_i,b_j$ as well as at $p = 0$.
Since the residue theorem says that the sum
of all residues is equal to $0$, one can conclude that
\begin{gather*}
  \left(\frac{\rd}{\rd b_i},\frac{\rd}{\rd b_j}\right)
  = 0 \qquad \text{for}\quad i \not= j .
\end{gather*}
By the same reasoning, one can conf\/irm that
\begin{gather*}
  \left(\frac{\rd}{\rd b_i},\frac{\rd}{\rd c_k}\right)
  = 0 \qquad \text{for}\quad k < N .
\end{gather*}
As regards the remaining cases, one can use the residue theorem
to rewrite the inner products as
\begin{gather*}
  \left(\frac{\rd}{\rd b_i},\frac{\rd}{\rd b_i}\right)
  = - \res_{p=b_i}\left[\frac{\kappa_i{}^2}{(p-b_i)^2}
        \left(\frac{\rd\log\lambda(p)}{\rd p}\right)^{-1}
        \frac{dp}{p^2}\right],\\
  \left(\frac{\rd}{\rd b_i},\frac{\rd}{\rd c_N}\right)
  = - \res_{p=0}\left[-\frac{\kappa_ip^{-N}}{p-b_i}
        \left(\frac{\rd\log\lambda(p)}{\rd p}\right)^{-1}
        \frac{dp}{p^2}\right].
\end{gather*}
In view of the local expression
\begin{gather*}
  \left(\frac{\rd\log\lambda(p)}{\rd p}\right)^{-1}
  = \frac{p-b_i}{\kappa_i} + O((p-b_i)^2
    \qquad\text{as}\quad p \to b_i
\end{gather*}
and
\begin{gather*}
  \left(\frac{\rd\log\lambda(p)}{\rd p}\right)^{-1}
  = - \frac{p^{N+1}}{Nc_N} + O\big(p^{N+2}\big)
    \qquad\text{as}\quad p \to 0 ,
\end{gather*}
one can readily calculate the residues
and conf\/irm the statement for the remaining cases.
\end{proof}

This lemma shows that $\log b_i$'s are part of f\/lat coordinates.
If one can further f\/ind a partial change of coordinates
$(c_1,\dots,c_N) \to (\qbar_0,\ldots,\qbar_{N-1})$
that is independent of $b_i$'s and f\/lat in themselves,
the new coordinate system
$\log b_1,\ldots,\log b_M$, $\qbar_0,\ldots,\qbar_{N-1}$
are totally f\/lat.  We shall show that
\begin{gather}\label{add2}
  \qbar_0 = \phi, \qquad
  \qbar_n = \res_{p=0}\left[\frac{\zbar(p)^n}{n}d\log p\right],\qquad
    n = 1,\ldots,N-1,
\end{gather}
give such coordinates.

To this end, we need to know some properties
of $\qbar_n$'s (which are def\/ined for $n \ge N$ as well
by the same residue formula).
Let $\pbar(\zeta)$ denote the inverse function
of $\zeta = \zbar(p)$.  It has a Laurent expansion of the form
\begin{gather*}
  \pbar(\zeta)
  = e^\phi\zeta^{-1}\big(1 + \ubar_1\zeta^{-1} + \cdots\big).
\end{gather*}
Therefore $\log\pbar(\zeta)$ is also well-def\/ined
as a series of the form
\begin{gather*}
  \log\pbar(\zeta)
  = - \log\zeta + \phi + \ubar_1\zeta^{-1} + \cdots.
\end{gather*}

\begin{lemma}
$\qbar_n$'s coincide with the coefficients of
the expansion of $\log\pbar(\zeta)$:
\begin{gather}
  \log\pbar(\zeta)
  = - \log\zeta + \qbar_0 + \qbar_1\zeta^{-1}
    + \cdots + \qbar_n\zeta^{-n} + \cdots.
\label{log(pbar)-expansion}
\end{gather}
\end{lemma}

\begin{proof}
One can rewrite the def\/inition \eqref{add2} of $\qbar_n$'s as
\begin{gather*}
  \qbar_n  = \res_{p=0}\left[\frac{\zbar(p)^n}{n}d\log p\right]
   = \res_{\zeta=\infty}\left[\frac{\zeta^n}{n}d\log\pbar(\zeta)\right] \\
  \hphantom{\qbar_n}{}
   = - \res_{\zeta=\infty}\left[\log\pbar(\zeta)
          d\left(\frac{\zeta^n}{n}\right)\right]
   = - \res_{\zeta=\infty}\left[\log\pbar(\zeta)\zeta^{n-1}d\zeta\right].
\end{gather*}
This implies that $\log\pbar(\zeta)$ has a Laurent expansion
as~(\ref{log(pbar)-expansion}) shows.
\end{proof}

\begin{lemma}
For $n = 1,\ldots,N-1$, $\qbar_n$ is a polynomial
of $c_{N-n},\ldots,c_{N-1}$ and $e^{-\phi}$ of the form
\begin{gather}
  \qbar_n = \frac{1}{N}e^{(n-N)\phi}c_{N-n}
      + \text{higher orders in $c_{N-n+1},\ldots,c_{N-1}$}
\label{qbar_n(c)}
\end{gather}
and $q_N$ is a function of $b_i$'s only
\begin{gather}
  \qbar_N = \frac{1}{N}\log\prod_{i=1}^M(-b_i)^{\kappa_i}.
\label{qbar_N(b)}
\end{gather}
\end{lemma}

\begin{proof}
Since $\zbar(p) = (\log\lambda(p))^{1/N}$ and $c_N = e^{N\phi}$,
one can calculate its $n$-th power as a Laurent series
of the form
\begin{gather*}
  \zbar(p)^n
  = \left(e^{N\phi}p^{-N} + c_{N-1}p^{1-N} + \cdots + c_1p^{-1}
      + \log\prod_{i=1}^M(-b_i) + O(p)\right)^{n/N}.
\end{gather*}
Extracting the $p^0$ term yields~(\ref{qbar_n(c)})
and~(\ref{qbar_N(b)}).
\end{proof}

(\ref{qbar_n(c)}) implies that the map $(c_1,\ldots,c_N)
\mapsto (\qbar_0,\ldots,\qbar_{N-1})$ is invertible.
We now choose $b_i$'s and $\qbar_0,\ldots,\qbar_{N-1}$
as a new coordinate system on the parameter space of $\lambda(p)$,
and consider $\lambda(p)$ to be a function of $p$
and these coordinates.

\begin{lemma}
The derivatives of $\log\lambda(p)$ with respect to
$\qbar_n$'s can be expressed as
\begin{gather}
  \frac{\rd\log\lambda(p)}{\rd\qbar_n}
  = \big({-}\zeta^{-n} + O\big(\zeta^{-N-1}\big)\big)\big|_{\zeta=\zbar(p)}
    \frac{d\log\lambda(p)}{d\log p}
\label{dlog(qbar)-lambda(II)}
\end{gather}
in a neighborhood of $p = 0$.
\end{lemma}

\begin{proof}
We can use the so called
``thermodynamic identity'' \cite{Dubrovin-2dtft}
\begin{gather*}
  \frac{\rd\log\lambda(p)}{\rd\qbar_n}d\log p
  = - \left.\frac{\rd\log\pbar(\zeta)}{\rd\qbar_n}
      \right|_{\zeta=\zbar(p)}d\log\lambda(p)
\end{gather*}
to rewrite the derivatives of $\log\lambda(p)$
with respect to $\qbar_n$'s as
\begin{gather*}
  \frac{\rd\log\lambda(p)}{\rd\qbar_n}
  = - \left.\frac{\rd\log\pbar(\zeta)}{\rd\qbar_n}
      \right|_{\zeta=\zbar(p)}\frac{d\log \lambda(p)}{d\log p}.
\end{gather*}
Let us recall the expansion~(\ref{log(pbar)-expansion})
of $\log\pbar(\zeta)$. In this expansion,
the coef\/f\/icients of $\zeta^0,\ldots,\zeta^{N-1}$
are $\qbar_n$'s themselves and, as~(\ref{qbar_N(b)}) shows,
the next leading coef\/f\/icient $q_N$ is a function of $b_i$'s only.
Consequently,
\begin{gather*}
  \frac{\rd\log\pbar(\zeta)}{\rd\qbar_n}
  = \zeta^{-n} + O\big(\zeta^{-N-1}\big).\tag*{\qed}
\end{gather*}
  \renewcommand{\qed}{}
\end{proof}

\begin{remark}
This is a place where the technical details slightly deviate
from the case of Ferguson and Strachan~\cite{FS06}.
As in their case, the f\/lat coordinates under construction
are a mixture of the two types of coordinates,
$b_i$'s and $\qbar_n$'s, presented in the Case~I.
There is, however, a delicate dif\/ference in the derivation
of the vital equalities~(\ref{dlog(qbar)-lambda(II)}).
In their case, they could derive these equalities
in a rather straightforward manner.  In our case,
we need a small piece of extra consideration
on the special structure of~$q_N$ as explained above.
\end{remark}

\begin{theorem}\label{thm:II-flat-coord}
The inner product of the derivatives in~$b_i$'s and~$\qbar_n$'s can be expressed as
\begin{gather*}
  \left(\frac{\rd}{\rd b_i},\frac{\rd}{\rd b_j}\right)
    = - \delta_{ij}\frac{\kappa_i}{b_i{}^2},\qquad
  \left(\frac{\rd}{\rd\qbar_m},\frac{\rd}{\rd\qbar_n}\right)
    = N\delta_{m+n,N},\qquad
  \left(\frac{\rd}{\rd b_i},\frac{\rd}{\rd\qbar_n}\right)
    = -\delta_{n0}\frac{\kappa_i}{b_i}.
\end{gather*}
In particular, $\log b_i$'s and $\qbar_n$'s are flat coordinates.
\end{theorem}

\begin{proof}
Since the expression
\begin{gather*}
  \frac{\rd\log\lambda(p)}{\rd b_i} = - \frac{\kappa_i}{p - b_i}
\end{gather*}
of derivatives with respect to $b_i$'s persists to be true,
the foregoing calculations of the inner products
of $\rd/\rd b_i$'s are also valid.  To consider
the inner products containing $\rd/\rd\qbar_n$'s,
we note the equality
\begin{gather*}
  \frac{\rd\log\lambda(p)}{\rd\qbar_n}
  = \sum_{k=1}^N\frac{\rd c_k}{\rd\qbar_n}p^{-k}
\end{gather*}
as well.  Thus the 1-forms in the expression
\begin{gather*}
  \left(\frac{\rd}{\rd\qbar_m},\frac{\rd}{\rd\qbar_n}\right)
  = \sum_{n=1}^{M+N}\res_{p=\gamma_n}
      \left[\frac{\rd\log\lambda(p)}{\rd\qbar_m}
        \frac{\rd\log\lambda(p)}{\rd\qbar_n}
        \left(\frac{\rd\log\lambda(p)}{\rd p}\right)^{-1}
      \frac{dp}{p^2}\right],\\
  \left(\frac{\rd}{\rd b_i},\frac{\rd}{\rd\qbar_n}\right)
  = \sum_{n=1}^{M+N}\res_{p=\gamma_n}
      \left[\frac{\rd\log\lambda(p)}{\rd b_i}
        \frac{\rd\log\lambda(p)}{\rd\qbar_n}
        \left(\frac{\rd\log\lambda(p)}{\rd p}\right)^{-1}
      \frac{dp}{p^2}\right]
\end{gather*}
of the inner products are rational,
and can have extra poles at $p = 0$ in addition
to the f\/irst order poles at $p = q_n$'s.
By the residue theorem, the sum over the residues
at $q_n$'s can be converted to the residues at $p = 0$:
\begin{gather*}
  \left(\frac{\rd}{\rd\qbar_m},\frac{\rd}{\rd\qbar_n}\right)
  = - \res_{p=0}
      \left[\frac{\rd\log\lambda(p)}{\rd\qbar_m}
        \frac{\rd\log\lambda(p)}{\rd\qbar_n}
        \left(\frac{\rd\log\lambda(p)}{\rd p}\right)^{-1}
      \frac{dp}{p^2}\right],\\
  \left(\frac{\rd}{\rd b_i},\frac{\rd}{\rd\qbar_n}\right)
  = - \res_{p=0}
      \left[- \frac{\kappa_i}{p-b_i}
        \frac{\rd\log\lambda(p)}{\rd\qbar_n}
        \left(\frac{\rd\log\lambda(p)}{\rd p}\right)^{-1}
      \frac{dp}{p^2}\right].
\end{gather*}
To evaluate the residues of $p = 0$,
let us recall~(\ref{dlog(qbar)-lambda(II)}).
The residues in the last equalities can be thereby evaluated as
\begin{gather*}
    \res_{p=0}\left[
        \frac{\rd\log\lambda(p)}{\rd\qbar_m}
        \frac{\rd\log\lambda(p)}{\rd\qbar_n}
        \left(\frac{\rd\log\lambda(p)}{\rd p}\right)^{-1}
        \frac{dp}{p^2}\right]\\
  \qquad{} = \res_{p=0}\big[
     \big({-}\zeta^{-m} + O\big(\zeta^{-N-1}\big)\big)
     \big({-}\zeta^{-n} + O\big(\zeta^{-N-1}\big)\big)\big|_{\zeta=\zbar(p)}
     d\log\lambda(p)\big]\\
  \qquad{} = \res_{\zeta=\infty}\big[
     \big({-}\zeta^{-m} + O\big(\zeta^{-N-1}\big)\big)
     \big({-}\zeta^{-n} + O\big(\zeta^{-N-1}\big)\big)
     N\zeta^{N-1}d\zeta\big]
   = - N\delta_{m+n,N}
\end{gather*}
and
\begin{gather*}
    \res_{p=0}\left[
        \frac{\rd\log\lambda(p)}{\rd b_i}
        \frac{\rd\log\lambda(p)}{\rd\qbar_n}
        \left(\frac{\rd\log\lambda(p)}{\rd p}\right)^{-1}
        \frac{dp}{p^2}\right]\\
  \qquad{} = \res_{p=0}\left[
     \left(-\frac{\kappa_i}{p-b_i}\right)
     \big({-}\zeta^{-n} + O\big(\zeta^{-N-1}\big)\big)\big|_{\zeta=\zbar(p)}
     d\log p\right]
   =   \delta_{n0}\frac{\kappa_i}{b_i}.
\end{gather*}
This completes the proof.
\end{proof}

\begin{remark}
As in the case of Ferguson and Strachan \cite{FS06},
we have been unable to f\/ind a dual Frobenius structure.
A naive idea will be to consider the same inner product
(\ref{<,>(I)}) and the cubic form (\ref{<,,>(I)})
as used for Dubrovin and Zhang's Landau--Ginzburg potential.
This, however does not work because the 1-forms
in the def\/inition of the inner product and the cubic form
have essential singularities at $p = 0$,
and one cannot use the residue theorem.
\end{remark}

\begin{remark}\label{rem:II-homogeneity}
Unlike the Landau--Ginzburg potential
of Ferguson and Strachan \cite{FS06},
our Landau--Ginzburg potential (\ref{lambda(II)})
has natural quasi-homogeneity.  We can use the vector f\/ield
\begin{gather*}
  E = \sum_{n=1}^{M+N}\lambda_n\frac{\rd}{\rd\lambda_n}
    = \frac{1}{\Mtilde}\sum_{i=1}^M b_i\frac{\rd}{\rd b_i}
      + \frac{1}{\Mtilde}\sum_{k=1}^N kc_k\frac{\rd}{\rd c_k}
\end{gather*}
as an Euler vector f\/ield.
\end{remark}

\section{Conclusion}\label{section6}

\looseness=-1
We have examined the two Landau--Ginzburg potentials
(\ref{lambda(I)}) and (\ref{lambda(II)}) and
the associated reductions of the dispersionless Toda hierarchy.
(\ref{lambda(I)}) is a generalization of Dubrovin and Zhang's
Landau--Ginzburg potential \cite{DZ98}, and contains
distinct examples such as the dispersionless
Ablowitz--Ladik hierarchy \cite{BCR11}.
(\ref{lambda(II)}) is a transcendental function,
and its logarithm resembles Ferguson and Strachan's
generalized waterbag model \cite{FS06}
of the dispersionless KP hierarchy.

We have observed that these quite dif\/ferent
 Landau--Ginzburg potentials have very similar features:
\begin{itemize}\itemsep=0pt
\item Consistency of the reduction can by conf\/irmed
by a direct method as shown in the proof of
Theorem~\ref{thm:reduced-Lax-eq}.
\item The radial L\"owner equations can be derived
in much the same way as shown
in the proof of Theorem~\ref{thm:Lowner-eq}.
Once these equations are obtained,
the generalized hodograph method works automatically
and in a unif\/ied way.
\item They have natural homogeneity.  This ensures f\/latness of
the third Egorov metric~(\ref{metric(hhat)}), hence
of the inner products~(\ref{(,)(I)}) and~(\ref{(,)(II)}).
Even the construction of f\/lat coordinates of these inner products
are partially similar as shown in the proof of
Theorem~\ref{thm:II-flat-coord}.
\end{itemize}

Meanwhile, we have encountered a rather complicated situation
in the problem of f\/latness of the f\/irst Egorov metric
(\ref{metric(h)}) (equivalently, the inner product
(\ref{<,>(I)})) in the cases other than Dubrovin and Zhang's
Landau--Ginzburg potential (\ref{DZ-lambda}).
It is well known that non-f\/latness of the Egorov metric implies
non-locality of an underlying Hamiltonian structure \cite{CLR09}.
Presumably, this issue should be considered case-by-case.
Of particular interest is the reduced Lax func\-tion~(\ref{lambda-dAL})
of the dispersionless Ablowitz--Ladik hierarchy, which is shown
to have a dual pair of Frobenius structures \cite{BCR11}.
Extending their result to other specializations
of (\ref{lambda(I)}) is an important open problem.

Our second Landau--Ginzburg potential (\ref{lambda(II)}),
too, raises many open problems.  Questions posed by Ferguson
and Strachan \cite{FS06} to the waterbag models
of the dispersionless KP hierarchy can be restated
for our model.  Moreover, the degenerate case (\ref{lambda(III)})
of this Landau--Ginzburg potential will have its own
special properties and applications.

\subsection*{Acknowledgements}

We thank the referees for many valuable comments.
This work is partly supported by JSPS Grants-in-Aid for
Scientif\/ic Research No.\ 21540218 and No.\ 22540186
from the Japan Society for the Promotion of Science.

\pdfbookmark[1]{References}{ref}
\LastPageEnding

\end{document}